\documentclass[aps,prl,twocolumn,superscriptaddress]{revtex4-2}

\usepackage[utf8]{inputenc}   % allows UTF-8 characters (ñ, á, é, etc.)
\usepackage[T1]{fontenc}      % proper font encoding for accented characters
\usepackage[english]{babel}

\usepackage{amsmath,amssymb,hyperref,graphicx,color}
\usepackage{mathtools,slashed}
\usepackage{caption}
\usepackage{subcaption}
\usepackage{braket}
\usepackage{ragged2e}
\usepackage{tensor}
\usepackage{comment}

\newcommand{\diff}{\text{d}}
\newcommand{\IAB}{I(A,B)}
\newcommand{\rh}{\hat{r}}

\newcommand{\GN}{G_{\rm N}}

\newcommand{\be}{\begin{equation}}
\newcommand{\ee}{\end{equation}}
\newcommand{\bea}{\begin{eqnarray}}
\newcommand{\eea}{\end{eqnarray}}

\def\({\left(}
\def\){\right)}
\def\[{\left[}
\def\]{\right]}

\DeclareMathOperator{\csch}{csch}
\DeclareMathOperator{\arcosh}{arcosh}

\definecolor{rossocorsa}{rgb}{0.83, 0.0, 0.0}
\definecolor{bleudefrance}{rgb}{0.19, 0.55, 0.91}
\hypersetup{
    colorlinks,
    citecolor=bleudefrance,
    filecolor=bleudefrance,
    linkcolor=rossocorsa,
    urlcolor=bleudefrance
}
\usepackage[normalem]{ulem}

\begin{document}
%%%%%%%
\noindent\flushright
{YITP-26-115} 
\noindent\flushleft\justifying
\vspace{-2cm}
%%%%%%%
\title{Monogamous quantum corrections to holographic mutual information}

\author{César A. Agón}
\affiliation{Universidad Politécnica de Cartagena, member of European University of Technology EUT+,\\
C/Dr Fleming S/N 30202 Cartagena, Spain}
\author{Javier Molina-Vilaplana}
\affiliation{Universidad Politécnica de Cartagena, member of European University of Technology EUT+,\\
C/Dr Fleming S/N 30202 Cartagena, Spain}
\author{Javier Moreno}
\affiliation{Center for Gravitational Physics and Quantum Information,\\
Yukawa Institute for Theoretical Physics, Kyoto University,\\
Kitashirakawa Oiwakecho, Sakyo-ku, Kyoto 606-8502, Japan}

\begin{abstract}
\noindent 
We construct an extensive model of bulk mutual information (MI) in anti-de Sitter spacetime and use it to describe the leading quantum corrections to the MI of certain holographic theories. The equivalence between bulk and boundary MI follows from the Faulkner--Lewkowycz--Maldacena prescription. We show that the extensivity of the bulk MI is sufficient for the quantum-corrected boundary entropy vector to satisfy monogamy of MI and all other holographic entropy cone inequalities. Our model is built from a covariantly conserved current-current kernel in AdS and is realized explicitly by a free massive Dirac fermion, whose boundary MI at long distances is that of a fermionic generalized free field (GFF). The construction also reproduces the expected short-distance divergence structure of a GFF, and the current-current correlator defines a positive and
monotonically decreasing function along the renormalization-group flow. We illustrate this explicitly for disjoint half-disks at the conformal fixed point of the model in AdS$_3$, where we present the exact result.
\end{abstract}

\maketitle
\noindent\textbf{Introduction}.
Mutual information (MI) between spatially disjoint regions is a finite and universal probe of correlations in quantum field theory (QFT). It is positive and monotone under inclusion, it provides an upper bound on connected correlators, and it encodes the operator content of the theory through its long-distance expansion~\cite{Casini:2006ws,Wolf:2007tdq,Cardy:2013nua}. However, this quantity is famously difficult to compute: explicit closed-form expressions are scarce, and thus there is little intuition about its geometric dependence \cite{Casini:2009vk,Arias:2018tmw,Abate:2026yzt}. An important exception is provided by QFTs admitting a holographic dual description through the anti-de Sitter/conformal field theory (AdS/CFT) correspondence~\cite{Maldacena:1997re,Gubser:1998bc,Witten:1998qj}, for which MI can be computed geometrically via the Ryu--Takayanagi (RT) prescription~\cite{Ryu:2006bv,Ryu:2006ef,Hubeny:2007xt,Nishioka:2009un,Lewkowycz:2013nqa,Rangamani:2016dms}. Interestingly, holographic entropies satisfy additional constraints known as the holographic entropy cone inequalities, among which monogamy plays a special role~\cite{Hayden:2011ag,Bao:2015bfa}---see also, e.g.,~\cite{Mirabi:2016elb,Cui:2018dyq,Agon:2018lwq,Hubeny:2018bri,HernandezCuenca:2019wgh,Czech:2019lps,Czech:2023xed,Grado-White:2024gtx,Grimaldi:2025jad}. However, the leading classical description of holographic MI is incomplete, as evidenced by the presence of a sharp transition that makes it vanish at sufficiently large separation. This would incorrectly suggest the absence of correlations beyond a certain scale. The resolution is provided by the Faulkner--Lewkowycz--Maldacena (FLM) prescription~\cite{Faulkner:2013ana}, which establishes that, at the first subleading order in the large-$N$ expansion, the holographic MI in the disconnected phase receives a contribution from the bulk MI associated with the dual entanglement wedges. Thus, a complete description of holographic MI must include the subleading ${\cal O}(1)$ terms.

A complementary approach, motivated by the difficulty of computing these subleading corrections directly, is to construct phenomenological models of MI that satisfy its known structural properties by construction. This program was initiated by Casini and Huerta, who introduced the extensive mutual information (EMI) model~\cite{Casini:2008wt}. With extensivity imposed, the EMI model is the unique solution to the MI constraints in a CFT, and remarkably, for $d=2$ it exactly reproduces the free-fermion result \cite{Casini:2005rm}. Although the EMI model is known to fail to capture the true MI of any local QFT for $d>2$, it remains a valuable toy model for exploring the geometric dependence of MI in higher dimensions, as it expresses the MI as a bilocal functional whose kernel depends only on the relative geometry of pairs of points~\cite{Casini:2015woa,Bueno:2015rda,Witczak-Krempa:2016jhc,Bueno:2019mex,Estienne:2021hbx,Agon:2021zvp,Bueno:2021fxb,Bueno:2023gey,Bueno:2026hku}.

Combining these perspectives, we construct an extensive functional for the bulk MI in rigid AdS, built, as in~\cite{Casini:2008wt}, from a covariantly conserved current-current kernel. As a first consequence, extensivity of the bulk MI implies monogamy of mutual information (MMI) for the leading quantum corrections to the MI of the corresponding boundary regions~\footnote{See \cite{Agon:2021lus} for a discussion on the OPE constraints on monogamous CFTs arising from the long-distance expansion of $I_3$}. Similar properties arise for geometric quantum corrections in double-holographic setups~\cite{Neuenfeld:2021bsb,Agon:2021tia}. We instantiate the construction with a bulk free massive Dirac fermion, whose current-current correlator has asymptotics that identify a fermionic generalized free field (GFF)~\cite{Duetsch:2002hc, Benedetti:2022aiw} as the lowest-dimensional boundary primary. The same correlator defines a positive, monotonically decreasing function along the renormalization group (RG) flow~\cite{Abate:2024nyh,Abate:2026apg}, which we use to identify the conformal fixed point of the model in AdS. A Weyl relation to Minkowski half-space~\cite{Takayanagi:2011zk,Fujita:2011fp} then yields a surface representation of the conformal MI, allowing us to infer the short-distance divergence structure expected for a boundary GFF. We illustrate this explicitly in AdS$_3$, where the MI can be computed exactly for disjoint half-disks.

\noindent\textbf{Monogamy and extensivity in MI}.
For two spatially disjoint regions $X$ and $Y$ in a QFT, the MI is $I_2(X,Y)\equiv I(X,Y) = S(X) + S(Y) - S({X\cup Y})$, where $S(X) = -{\rm Tr}\!\left[\rho_X \log \rho_X\right]$ is the von Neumann entropy of the reduced density matrix $\rho_X = {\rm
Tr}_{\bar X}|\Omega\rangle\langle\Omega|$, with $|\Omega\rangle$ the QFT vacuum and the bar denoting the spatial complement. Although the individual entropies are UV-divergent~\footnote{For renormalized entanglement entropy in a holographic context, see~\cite{Taylor:2016aoi,Anastasiou:2017xjr,Taylor:2020uwf,Anastasiou:2020smm,Anastasiou:2021swo,Anastasiou:2022ljq,Anastasiou:2024rxe,Anastasiou:2026bbf}.}, their divergences cancel in $I(X,Y)$, yielding a finite continuum quantity~\cite{Witten:2018zxz}. Subadditivity of the entropy implies non-negativity of the MI, while strong subadditivity implies monotonicity under inclusion~\cite{Lieb:1973cp},
\begin{equation}\label{eq:subpos}
I(X,Y\cup Z)\geq I(X,Y)\geq0\,.
\end{equation}
The MI is also invariant under the spacetime symmetries of the vacuum---and under conformal transformations in a CFT---and bounds connected correlators~\cite{Casini:2006ws,Wolf:2007tdq}.

For infinitesimally separated regions, the MI diverges proportionally to the area of their common boundary, $I(X,\bar{X}_{\varepsilon})\sim \varepsilon^{-(d-2)}$, where $\bar{X}_{\varepsilon}=\overline{X+\varepsilon}$ and $\varepsilon$ denotes their separation. The coefficients of the universal terms in this short-distance expansion play a fundamental role in characterizing RG monotones in QFT~\cite{Casini:2008wt,Casini:2015woa}.

MI may also satisfy additional properties, such as
\begin{align}
\text{monogamy: } &I_3(X,Y,Z)\leq 0\,,\label{eq:monog}\\
\text{extensivity: } &I_3(X,Y,Z)= 0\,,
\label{eq:exten}
\end{align}
where $I_3(X,Y,Z)\equiv I(X,Y) + I(X,Z) - I(X,Y\cup Z)$ is known as the tripartite information.

A canonical realization of monogamous MI is provided by holography. In holographic theories, the entropy of a boundary region $A_\partial$ is given semiclassically by~\cite{Ryu:2006bv,Ryu:2006ef,Hubeny:2007xt,Nishioka:2009un,Lewkowycz:2013nqa,Faulkner:2013ana,Rangamani:2016dms}
\begin{equation}
S^{\rm holo}(A_\partial)=\frac{\mathrm{Area}(\Gamma_{A})}{4\GN}
+S^{\rm b}(A)+\mathcal{O}(\GN)\,,
\label{eq:rt}
\end{equation}
where $\Gamma_{A}$ is the minimal bulk surface homologous to $A_\partial$, $A$ is its entanglement wedge, and $S^{\rm b}$ is the bulk entropy. At leading order, the geometric structure of RT entropies implies monogamy of MI~\eqref{eq:monog}~\cite{Hayden:2011ag}. It also produces a sharp connected-to-disconnected transition: for sufficiently separated $A_\partial$ and $B_\partial$, the classical MI vanishes \cite{Headrick:2010zt}. In this disconnected phase the area contributions cancel, and the leading nonvanishing term is instead the MI of bulk quantum fields,
$I^{\rm holo}(A_\partial,B_\partial)=I^{\rm b}(A,B)+\mathcal{O}(\GN)$~\cite{Faulkner:2013ana,Agon:2015ftl}.

On the other hand, Casini and Huerta introduced the flat-space EMI model based on extensivity~\cite{Casini:2008wt}. Imposing causality, Lorentz invariance, and additivity over subregions of a Cauchy slice, the MI can be written as the bilocal functional
\begin{equation} 
    I(X,Y) = \kappa \int_X \int_Y \diff \sigma_X\,\diff\sigma_Y\, n_X^{\mu} n_Y^{\nu}J_{\mu \nu}\, ,
    \label{eq:mi_emi}
\end{equation}
where $\kappa$ is a normalization constant, $\diff\sigma_{X}\equiv\sqrt{h_{X}}\,\diff^{d-1}x_X$ denotes the induced volume element on the region $X$ of a Cauchy slice, $n_{X}^{\mu}\equiv n^{\mu}(x_X)$ is its future-directed unit normal, and the kernel $J_{\mu\nu}\equiv J_{\mu\nu}(x_X,x_Y)$ is conserved with respect to both indices,
\begin{equation}\label{eq:conseq}
\nabla _{\mu}^{X}J^{\mu \nu}=0\,,\qquad \nabla _{\nu}^{Y}J^{\mu \nu}=0\,.
\end{equation}
The kernel can take the form of a conserved current-current correlator, $J_{\mu\nu}(x_X,x_Y) \equiv\langle J_\mu(x_X)J_\nu(x_Y)\rangle$, which, by Lorentz invariance, can be written as 
\begin{equation}
\langle J_\mu(x_X)J_\nu(x_Y)\rangle
=\frac{G(r)\rh_\mu\rh_\nu-F(r)\eta_{\mu\nu}}{r^{2(d-1)}}\,,
\label{eq:j_corr}
\end{equation}
where $r\equiv|x_X-x_Y|$ is the distance between the points, $\rh_\mu\equiv(x_X-x_Y)_\mu/r$ is the unit separation vector, and $\eta_{\mu\nu}$ is the flat-space metric. 

The monotonicity of the EMI model is guaranteed by requiring positivity of the kernel for every pair of points on the integration hypersurfaces, $n_X^{\mu}\, n_Y^{\nu}J_{\mu \nu}(x_X,x_Y) \geq 0$, for all future-directed timelike vectors $n_X^{\mu}$ and $n_Y^{\nu}$. This requirement translates into the conditions
\begin{equation}
    2F(r) \geq G(r)\geq 0\, ,
    \label{eq:monotonicity}
\end{equation}
and, in a CFT,
\begin{equation}\label{eq:GFk}
G(r)=2F(r)=2k_J\,,
\end{equation}
with $k_J$ a constant representing the current central charge. 

\begin{figure}
\centering
\includegraphics[width=1\linewidth]{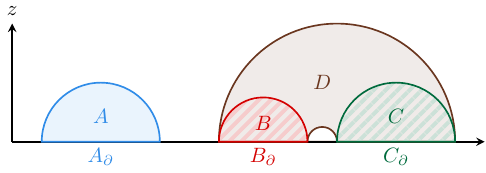}
\caption{\justifying Schematic representation of the entanglement wedges associated with the boundary regions $A_\partial$, $B_\partial$, and $C_\partial$. We denote $D\equiv{\rm EW}(B_\partial\cup C_\partial)$.}
\label{fig:fig1}
\end{figure}

\noindent\textbf{Bulk extensivity implies boundary monogamy and the holographic entropy cone inequalities}. Bulk MMI has been shown to imply boundary MMI within the quantum extremal surface (QES) prescription~\cite{Akers:2021lms}. In the disconnected RT phase considered here, bulk extensivity allows this implication to be seen directly at the level of the leading quantum correction.
Consider three disjoint boundary regions $A_\partial$, $B_\partial$, and $C_\partial$, with corresponding (disjoint) entanglement wedges  ({\rm EW}) $A$, $B$, and $C$, and define $D\equiv {\rm EW}(B_\partial\cup C_\partial)$~\cite{Faulkner:2013ana}---see Fig.~\ref{fig:fig1}. In the disconnected RT phase, the classical contribution to the MI vanishes, so that
$I^{\rm holo}(A_\partial,B_\partial\cup C_\partial)
=I^{\rm b}(A,D)+\mathcal{O}(\GN)$. Entanglement-wedge nesting implies $B,C\subseteq D$~\cite{Wall:2012uf}, and hence
$D=(B\cup C)\cup[D\backslash(B\cup C)]$.
Bulk extensivity then gives
$I^{\rm b}(A,D)
=I^{\rm b}(A,B)+I^{\rm b}(A,C)+I^{\rm b}\left(A,D\backslash(B\cup C)\right)$.
Since the last term is non-negative and the same disconnected-phase relation holds for the pairs $(A_\partial,B_\partial)$ and $(A_\partial,C_\partial)$, we obtain
\begin{equation}\label{eq:IbIholo}
I^{\rm holo}(A_\partial,B_\partial\cup C_\partial)
\!\geq
I^{\rm holo}(A_\partial,B_\partial)
+
I^{\rm holo}(A_\partial,C_\partial)\,,
\end{equation}
up to $\mathcal O(G_N)$ corrections.
Thus, an extensive model of bulk MI in AdS gives rise to a monogamous model on the boundary.

More generally,~\cite{Akers:2021lms} showed that, for $N$ regions, if the bulk entropy vector in the usual entropy basis, $
\mathbf S^{\rm b}
=
\left(
S^{\rm b}(A_i),
S^{\rm b}(A_i\cup A_j),
\ldots,
S^{\rm b}(A_1\cup\cdots\cup A_N)
\right)
$, with the indices running over all distinct subsets of ${1,\ldots,N}$, satisfies the holographic entropy cone (HEC) inequalities, $\mathbf S^{\rm b}\in{\rm HEC}$, then the corresponding boundary QES entropy vector also lies in the HEC,
\begin{equation}\label{eq:Akersetal}
\mathbf S^{\rm b}\in{\rm HEC}
\Longrightarrow
\mathbf S^{\rm holo}\in{\rm HEC}\,.
\end{equation}
Here the $A_i$ denote the bulk regions into which the candidate RT surfaces partition the bulk Cauchy slice, so that every entanglement wedge is a union of them; extensivity is imposed on these regions. Employing this result, we now show that positivity and extensivity of the bulk MI provide a sufficient condition for $\mathbf S^{\rm holo}\in{\rm HEC}$. To see this, consider the bulk entropy vector expressed in the multipartite-information basis,
$\mathbf I^{\rm b}
=
\left(
{S^{\rm b}(A_i)},
{I^{\rm b}(A_i,A_j)},
{I_3^{\rm b}(A_i,A_j,A_k)},
\ldots,
I_N^{\rm b}(A_1,\ldots,A_N)
\right).$
Combining extensivity~\eqref{eq:exten} and the recursive identity $I_n(A_1,\ldots,A_n)=I_{n-1}(A_1,\ldots,A_{n-1})
    +I_{n-1}(A_1,\ldots,A_{n-2},A_n)
    -I_{n-1}(A_1,\ldots,A_{n-1}\cup A_n)$~\cite{10.1098/rsif.2007.1079,Alishahiha:2014jxa}, we inductively find that
\begin{equation}
I_n^{\rm b}=0\,,\qquad n\geq3\,.
\label{eq:multipartite-vanish}
\end{equation}
All nonredundant holographic entropy inequalities other than subadditivity~\eqref{eq:subpos}  can be expressed in the multipartite-information basis without one- or two-party terms~\cite{He:2020xuo,Hernandez-Cuenca:2023iqh,Hubeny:2026vlk}. Schematically, any such inequality takes the form 
$Q_{\rm HEC}
=
\sum_{n=3}^{N}
\sum_{i_1<\ldots<i_n}
c_{i_1\ldots i_n}
I_n^{\rm b}(A_{i_1},\ldots,A_{i_n})
\geq0$,
for some coefficients $c_{i_1\cdots i_n}$. Equation~\eqref{eq:multipartite-vanish} therefore sets every term in $Q_{\rm HEC}$ to zero, so that all these inequalities are saturated. Subadditivity, which is the remaining independent constraint, follows from the positivity of the MI~\eqref{eq:subpos}. Hence $\mathbf S^{\rm b}\in{\rm HEC}$, and applying~\eqref{eq:Akersetal} gives
\begin{equation}
\text{positive and extensive }I^{\rm b}
\Longrightarrow
\mathbf S^{\rm holo}\in{\rm HEC}\,.
\end{equation}
Therefore, positive and exactly extensive bulk MI is sufficient for the quantum-corrected boundary entropy vector to satisfy all holographic entropy inequalities.

\noindent\textbf{Extensive model of MI in AdS}.
Motivated by the direct bulk-to-boundary relation~\eqref{eq:IbIholo}, we now explore extensivity of MI in AdS. The bilocal EMI formulation in \eqref{eq:mi_emi} can be generalized to maximally symmetric spacetimes, and the current-current correlator of a free fermion in AdS provides an admissible kernel for it, yielding a family of extensive models parametrized by the fermion mass.
Throughout, $d$ denotes the dimension of the spacetime on which the QFT lives; for a QFT in AdS$_d$ the boundary theory is $(d-1)$-dimensional.

In AdS, we can write the current-current correlator, analogous to \eqref{eq:j_corr} in flat space, as
\begin{equation}
J_{\mu \nu'}(x,x') = -\frac{G(\ell)t_\mu t_{\nu'}+F(\ell)g_{\mu \nu'}}{[2L\sinh\left(\ell/2L\right)]^{2(d-1)}}\, ,
\label{eq:j_corr_ads}
\end{equation}
where $L$ is the radius of AdS, and $\ell\equiv\ell(x,x')=\int_\gamma\diff s$ is the geodesic distance between $x$ and $x'$ along the geodesic $\gamma$.
 Here $t_{\mu}=\nabla_{\mu}\ell$ and $t_{\nu'}=\nabla_{\nu'}\ell$ are the tangent vectors at $x$ and $x'$, respectively, while $g_{\mu\nu'}(x,x')$ is the AdS parallel propagator~\cite{Allen:1985wd,Allen:1986qj,Kawano:1999au}. 

Defining $k(\ell)\equiv G(\ell)-F(\ell)$, the conservation of $J_{\mu \nu'}(x,x')$, together with inequality~\eqref{eq:monotonicity}, implies
\begin{equation}
     k'(\ell)=-\frac{(d-1)\left[2F(\ell)-G(\ell)\right]}{L \sinh(\ell/L)} \leq 0\,. 
\end{equation}
 Thus, $k(\ell)$ is a monotonically decreasing function. In the short-distance limit $\ell\to0$, the conformal condition $G=2F$ implies that $k(\ell)$ approaches $k_J$, identifying the AdS analogue of the conformal current-current correlator \eqref{eq:GFk} with $G(\ell)=2F(\ell)=2k_J$.

Remarkably, we can write the kernel in a form that explicitly solves the conservation equation \eqref{eq:conseq} in AdS as
\begin{equation}
 J\indices{_{\mu}^{\nu'}}(x,x')
   = -\nabla^\alpha \nabla_{\beta'}
     \left(g\indices{^{[\nu'}_\mu} g\indices{^{\beta']}_{\alpha}} H\right),
\end{equation}
where $g_{\alpha[\beta'}g_{\gamma]\delta'}=g_{\alpha\beta'}g_{\gamma\delta'}-g_{\alpha\gamma}g_{\beta'\delta'}$ and $H\equiv H(\ell)$ satisfies the first-order equation
\begin{equation}
H'
+\frac{(d-2)\tanh\left(\frac{\ell}{2L}\right)}{L} H
=
\frac{k(\ell)\, L\sinh\left(\frac{\ell}{L}\right)}
{(d-1)
\left[2L\sinh\left(\frac{\ell}{2L}\right)\right]^{2(d-1)}}
\,.
\label{eq:diff_H}
\end{equation}
As in flat space~\cite{Casini:2008wt}, this allows us to write the EMI functional as a surface integral over the boundaries of the entangling regions $A$ and $B$, equivalent to \eqref{eq:mi_emi}:
\begin{equation}\label{eq:emi_flat}
    \IAB= -\kappa\, \int_{\partial A}\int_{\partial B}\diff\sigma_A\, \diff\sigma_B\, \mathcal{T}_{AB}\, H(\ell)\, ,
\end{equation}
with
\begin{equation}
    \mathcal{T}_{AB}=(n_A\cdot n'_B)(\tilde{n}_A \cdot \tilde{n}'_B) -(\tilde{n}_A \cdot n'_B)(n_A \cdot \tilde{n}'_B)\, ,
    \label{eq:tensor_emi}
\end{equation}
where, for $X=A,B$, $n_X$ and $\tilde n_X$ denote a pair of mutually orthogonal unit vectors normal to the codimension-two surface $\partial X$, with $n_X$ future-directed and $\tilde{n}_X$ pointing outward from $X$, and the prime indicates that the inner products are evaluated using the parallel propagator $g_{\mu \nu'}$---see the Supplemental Material for additional details on the derivation of the above formulas.

\noindent\textbf{EMI in AdS for a massive fermion}. For concreteness, we build an explicit model from the current of a free massive Dirac fermion of mass $m$ in AdS$_d$, $J_\mu(x)\equiv\bar\Psi(x)\gamma_\mu\Psi(x)$,
associated, through the standard field-operator map, with a boundary fermionic operator of scaling dimension $\Delta = (d-1)/2 + \nu$ with $\nu=mL$~\cite{Henningson:1998cd,Mueck:1998iz,Henneaux:1998ch}. Its two-point function takes the form~\cite{Allen:1985wd,Allen:1986qj,Kawano:1999au}
\begin{equation}
    \langle \Psi(x)\bar{\Psi}(x')\rangle = \left(a + b\, \gamma^{\mu} t_\mu\right)\Lambda(x,x')\, ,
\end{equation}
where $\gamma^\mu$ are the Dirac matrices and $\Lambda(x,x')$ is the parallel spinor propagator along the geodesic connecting $x$ and $x'$, and $a\equiv a(\ell)$, $b\equiv b(\ell)$ are scalar functions of the geodesic distance $\ell$. Both are fixed, up to overall normalization, by the Dirac equation and standard boundary conditions; their explicit hypergeometric form is given in \cite{Mueck:1999efk}.

Wick-contracting $\langle J_\mu(x)J_{\nu'}(x')\rangle$ and tracing over spinor indices---which produces the overall factor $2^{\lfloor d/2\rfloor}$ from the dimension of the Dirac spinor representation~\cite{Freedman:2012zz}---reproduces precisely the tensor structure of~\eqref{eq:j_corr_ads}, with
\begin{equation}
    F=\frac{2^{\lfloor d/2\rfloor}(a^2+b^2)}{[2L\sinh(\ell/2L)]^{2(1-d)}}\,, \quad G= \frac{2^{\lfloor d/2\rfloor+1} b^2}{[2L\sinh(\ell/2L)]^{2(1-d)}}\,.
\end{equation}

The explicit forms of $a\equiv a(\ell)$ and $b\equiv b(\ell)$, expressed in terms of $\zeta=\cosh^2(\ell/2L)$, are
\begin{align}
a&=\frac{\lambda}{\zeta^\Delta}\,{}_2F_1\left(\frac{d}2+\nu , \nu, 1+2\nu;\frac{1}{\zeta}\right)\,,\\
b&
=
\lambda\,
\frac{\sqrt{\zeta-1}}{\zeta^{\Delta+\frac12}}
\,{}_2F_1\!\left(
\frac{d}{2}+\nu,\,
1+\nu,\,
1+2\nu;\,
\frac{1}{\zeta}
\right)\,,
\end{align}
where
$\lambda$ is a constant---see the Supplemental Material. These expressions automatically satisfy the model's monotonicity conditions: $2F-G\propto a^2$ and $G>0$, guaranteeing positivity and monotonicity of the resulting EMI. Moreover, $k(\ell)$, explicitly given by
\begin{equation}    k(\ell)=\frac{2^{\lfloor d/2\rfloor}\, (b^2-a^2)}{[2L\sinh(\ell/2L)]^{2(1-d)}}\, ,
    \label{eq:kfunction}
\end{equation}
is non-negative for all $\ell$, satisfies $k'\leq 0$ because $2F-G\geq0$, and tends to zero as $\ell \to \infty$---see the Supplemental Material. It can therefore be interpreted as an RG monotone tracking the effective degrees of freedom ~\cite{Zamolodchikov:1986gt,Cardy:1988cwa,Klebanov:2011gs,Komargodski:2011vj,Casini:2012ei} coupled to the conserved U$(1)$ current in AdS, with the flow parametrized by $\ell$ and $\ell/L$ playing the role of an RG scale. This is in the spirit of the entropic $c$- and $F$-theorems for QFTs in flat space~\cite{Casini:2004bw,Casini:2012ei,Casini:2017vbe}, and of the recent program of monotonicity theorems for QFTs in rigid AdS~\cite{Abate:2026apg}---see also~\cite{Abate:2024nyh,Bason:2025sxb}. 

\noindent \textbf{Long- and short-distance structure}.
We now analyze the asymptotic behavior of the AdS EMI in Poincaré coordinates, $x=(z,\mathbf{x})$, for which $\ell=L\arcosh\left[1+(x-x')^2/(2zz')\right]$. For $X=A,B$, we choose the integration region $X$ to extend trivially from the RT surface $\Gamma_X$ to the AdS boundary, with $n_X^z=0$, and evaluate~\eqref{eq:mi_emi}.
\begin{figure}
    \centering
    \includegraphics[width=1\linewidth]{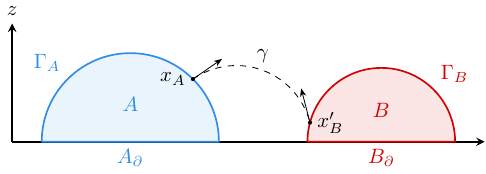}
    \caption{\justifying Two points $x_A$ and $x_B'$, expressed in Poincar\'e coordinates, lie on $\partial A=\Gamma_A\cup A_\partial$ and $\partial B=\Gamma_B\cup B_\partial$, respectively. They are connected by the geodesic $\gamma$.}
    \label{fig:fig2}
\end{figure}
Consider two compact regions $A$ and $B$ such that $|x_A-x_B|\gg z_A,z_B$ for all points in $A$ and $B$. We call this the long-distance limit. In that case, 
 the explicit solution gives $G\approx F$, and evaluating~\eqref{eq:mi_emi} using~\eqref{eq:j_corr_ads} we obtain 
\begin{equation}\label{Long-dist-mutual}
I(A,B)\sim 
\frac{\kappa\, k_\infty\, C_A C_B\left[2({\bf n}_A \cdot {\bf \hat{r}}) ({\bf n}_B \cdot {\bf \hat{r}})-{\bf n}_A \cdot {\bf n}_B\right]}{|\mathbf{x}_A-\mathbf{x}_B|^{4\Delta}}  \,,
\end{equation}
where ${\bf \hat{r}}$ is the unit vector along the boundary separation between two points in the interior of the boundary regions $A_{\partial}$ and $B_{\partial}$ denoted by $\mathbf{x}_A$ and $\mathbf{x}_B$, and $C_X=\int_{X} \diff^{d-1}x_X z_X^{2\Delta-(d-1)}$. Here $k_\infty=2^{\lfloor d/2\rfloor-1}\Gamma^2(d/2+\nu)/[\pi^{d-1}\Gamma^2(1/2+\nu)]$ fixes the normalization of the long-distance falloff of the kernel---see the Supplemental Material. Unlike the UV value $k_J$, $k_\infty$ depends on $\nu$. For fixed boundary regions and an appropriate choice of $\kappa$, this reproduces the leading long-distance MI of a CFT whose lowest-dimensional primary is a fermion of dimension $\Delta$~\cite{Casini:2021raa,Benedetti:2022aiw,Agon:2024zae}. The coefficients $C_X$ thus provide a prediction for the shape dependence of this contribution, which can be tested, for instance, in free theories.

We next turn to the short-distance regime, for which it is convenient to consider the conformal fixed point of the model. Rather than taking the conformal limit of the massive Dirac solution directly \footnote{The $m\to0$ limit admits two solutions differing by the sign of $a$. The condition $k(\ell)=k_J$ selects their average, which retains only the $b$ contribution. This defines the conformal point of the model; it is not the massless limit of either solution separately.}, we set $G=2F=2k_J$, so that $k(\ell)=k_J$, and solve~\eqref{eq:diff_H}.
The solution for $H(\ell)$ is expressed in terms of a hypergeometric function, making \eqref{eq:emi_flat} difficult to evaluate---see the Supplemental Material. Instead, it is more convenient to start from~\eqref{eq:mi_emi} and use its Weyl equivalence with Minkowski half-space~\cite{Takayanagi:2011zk,Fujita:2011fp}. In Poincaré coordinates, the conformal current-current correlator~\eqref{eq:j_corr_ads} is related to its flat-space counterpart by a local rescaling following from the AdS bitensor identities~\cite{DHoker:1999bve}, while the induced measures transform with the compensating factors---see the Supplemental Material. Consequently, the EMI functional~\eqref{eq:mi_emi} in AdS becomes identical to its counterpart in the Minkowski half-space. Thus, starting with the flat-space version of~\eqref{eq:mi_emi} and using Stokes' theorem, we arrive at a flat-space boundary integral representation of \eqref{eq:emi_flat}, namely 
\begin{align}\label{eq:emi_flat-bdy}
    \IAB= \frac{\kappa\, k_J}{2(d-1)(d-2)}\, \int_{\partial A}\int_{\partial B}
          \frac{\diff\sigma_A\, \diff \sigma_B\,\mathcal{T}_{AB}}
               {|x_A-x_B|^{2(d-2)}}\, ,
\end{align}
where the inner products in \eqref{eq:tensor_emi} are taken with the Minkowski metric. 

The boundary of each bulk region has two components, $\partial A=A_\partial\cup\Gamma_A$ and $\partial B=B_\partial\cup\Gamma_B$---see Fig.~\ref{fig:fig2}. Denoting by $\mathcal I_{XY}$ the contribution to~\eqref{eq:emi_flat-bdy} obtained by restricting its integration domains to $X$ and $Y$, respectively, we have
\begin{equation}
\IAB=
\mathcal I_{A_\partial B_\partial}
+\mathcal I_{\Gamma_A\Gamma_B}
+\mathcal I_{A_\partial\Gamma_B}
+\mathcal I_{\Gamma_A B_\partial}\,.
\end{equation}
 For disjoint spheres, the mixed terms cancel against each other. The term $\mathcal I_{A_\partial B_\partial}$, whose surfaces lie on the $(d-1)$-dimensional boundary, has exactly the same power-law divergences as flat-space EMI in $(d-1)$ dimensions; its explicit formula differs only by a mild angular dependence. On the other hand, $\mathcal I_{\Gamma_A\Gamma_B}$ corresponds to the $d$-dimensional EMI applied to half-hemispheres, and therefore has the same power-law divergences as those for full spheres, up to boundary terms. These boundary terms can only introduce additional divergences of the same type as those present in $\mathcal I_{A_\partial B_\partial}$. Therefore, we conclude that  
\begin{equation}\label{eq:sle}
\IAB\sim c_{d-2} \frac{R^{d-2}}{\varepsilon^{d-2}}+c_{d-3} \frac{R^{d-3}}{\varepsilon^{d-3}}+\ldots+c_0 \log\frac{R}{\varepsilon}\,,
\end{equation}
where the coefficients $c_i$ are universal, regulator-independent data of the theory. This is precisely the short-distance structure expected for a GFF~\cite{Benedetti:2022aiw}. In the massive Dirac realization, the coefficients become functions of $d$ and $\Delta$. 

As an explicit check, we evaluate the conformal EMI in AdS$_3$ for a half-disk $B$ of radius $R_B$ and the exterior $A$ of a concentric half-disk of radius $R_A=R_B+\varepsilon$, and define their average radius as $\bar R\equiv(R_A+R_B)/2$. The integrals in~\eqref{eq:emi_flat-bdy} can be performed exactly. Defining $\eta\equiv R_B/R_A$, we obtain
\begin{align}
I(A,B)=&
\frac{\kappa k_J\pi^2\eta^2}{4(1-\eta^2)}
+\frac{\kappa k_J}{2}\left(\frac{1+\eta^2}{1-\eta^2}\right)
\left[{\rm Li}_2(\eta)-{\rm Li}_2(-\eta)\right]\notag\\
&+\frac{\kappa k_J}{2}\log\(\frac{1-\eta}{1+\eta}\)\,.
\end{align}
The concentric configuration is conformally equivalent to a pair of disjoint half-disks of equal radius $R$ whose centers are separated by a distance $s$, with $s/R=(1+\eta)/\sqrt{\eta}$. Thus, $\eta\to0$ corresponds to the long-distance regime $s\gg R$, whereas $\eta\to1^{-}$ gives the short-distance regime $\varepsilon=R_A-R_B\ll\bar R$. The two limits are
\begin{equation}\label{eq:slethree}
I(A,B)\sim
\begin{cases}
\displaystyle
\frac{\kappa k_J\pi^2\, R^{4}}{s^{4}}\,,
& s\gg R\,,\\[1ex]
\displaystyle
\frac{\kappa k_J\pi^2\, \bar R}{4\varepsilon}\,
-\kappa k_J\log\frac{\bar R}{\varepsilon}
+\mathcal O(1)\,,
& \varepsilon\ll\bar R\,.
\end{cases}
\end{equation}
The long-distance result has the expected $s^{-4}$ falloff for a fermionic GFF with $\Delta=1$ and is one quarter of the complete-disk result in flat space~\cite{Casini:2021raa,Benedetti:2022aiw,Agon:2024zae}. At the conformal point, $k(\ell)=k_J$ for all $\ell$, so the long-distance coefficient $k_\infty$ in \eqref{Long-dist-mutual} reduces to $k_J$. In the short-distance limit, the leading area-law divergence agrees with the flat-space EMI result in $d=3$~\cite{Casini:2015woa}, while the logarithmic term realizes the characteristic GFF divergence structure~\cite{Benedetti:2022aiw}. Taking directly from~\eqref{eq:slethree}, the universal ``strip'' and logarithmic coefficients are, respectively, $c_1=\kappa k_J\pi^2/4$ and $c_0=-\kappa k_J$. 

\noindent\textbf{Discussion}. 
We have developed an extensive model for bulk MI in AdS and used it to characterize the leading quantum corrections to holographic MI. Its extensivity ensures monogamy of the boundary MI and, more generally, places the quantum-corrected boundary entropy vector inside the holographic entropy cone. The model is formulated in terms of a conserved current-current kernel and admits an explicit realization through a free massive Dirac fermion. This realization reproduces the expected fermionic generalized-free-field behavior at long distances and the corresponding short-distance divergence structure, including the logarithmic term. 

Several directions deserve further study. An immediate problem is to determine the universal coefficients $c_i$ entering the short-distance expansion~\eqref{eq:sle} in arbitrary dimension and to understand their dependence on $d$ and $\Delta$. It would also be interesting to test the predicted shape dependence of the long-distance contribution beyond spherical regions and to compare the exact results of the model for disjoint spheres with recent analytic approximations to generalized-free-field MI in higher dimensions~\cite{Agon:2024xvs,Agon:2026jjh}. More generally, one may seek other realizations of the extensive kernel, including different free bulk fields and interacting theories, and investigate whether the associated RG monotones admit a broader field-theoretic interpretation. We plan to address these directions in the future~\cite{Agon:XX}.

A further direction is to extend the construction to QFTs in de Sitter (dS) space. Recent results on RG irreversibility in fixed dS backgrounds~\cite{Abate:2024nyh}, together with analogous developments in rigid AdS~\cite{Abate:2026apg}, motivate comparing their entropic monotones with the correlator-based monotone found here. An AdS--dS continuation may also provide a simple setting in which to explore implications for dS/CFT~\cite{Strominger:2001pn,Strominger:2001gp,Maldacena:2002vr,Anninos:2012qw} and possible connections to cosmological entropy and pseudoentropy~\cite{Nakata:2020luh,Mollabashi:2020yie,Mollabashi:2021xsd,Doi:2023zaf,Narayan:2023ebn,Anastasiou:2025rvz,Anastasiou:2026bbf,Fujiki:2026kca}.

\noindent\emph{Acknowledgments}. We thank Horacio Casini, Matthew Headrick, Tadashi Takayanagi, and Nicolò Zenoni for useful discussions. C.A. is supported by the Saavedra Fajardo program (22824/SF/24) of the Fundación Séneca de la Región de Murcia and by the Spanish MINECO grant PID\-2024\-155685\-NB\-C22. J.M.-V. acknowledges support from the same MINECO grant and from the Fundación Séneca de la Región de Murcia (FSRM/\-10.13039/\-100007801, 22581/\-PI/\-24). J.M. is supported by the European Union's Horizon Europe research and innovation programme under the Marie Sk\l{}o\-dow\-ska\-Curie
No.~101\-202\-710 (QUBITTO).

\onecolumngrid

\bibliographystyle{apsrev4-2}
\bibliography{EE.bib}
\newpage
\vspace{1cm} 
\begin{center}  
{\Large\bf Supplemental Material} 
\end{center} 

\section{Current-current correlator in AdS}
In a maximally symmetric spacetime such as AdS, two-point functions of local scalar operators are functions of the geodesic distance $\ell(x,x')\equiv\int_x^{x'} \diff s$ between the points. If the local operators carry Lorentz indices, the associated correlator can be expanded in terms of a basis of bilocal tensors constructed from the parallel propagator bitensor $g_{a b'}(x,x')$ and the unit vectors tangent to the geodesic at the endpoints $t_a(x,x')$ and $t_{b'}(x,x')$~ \cite{Allen:1985wd}. Here $\{a, b, \cdots\}$ are tangent-space indices associated with $x$, while $\{a', b', \cdots\}$ are associated with $x'$. The tangent vectors are given by 
\begin{equation}\label{normals}
t_a(x,x')=\nabla_a \ell(x,x')\,, \qquad {\rm and}\qquad  t_{b'}(x,x')=\nabla_{b'} \ell(x,x')\,.
\end{equation}
Sometimes we will omit writing the base points $x$, $x'$ to avoid clutter. The following tensor relations will be useful: 
\begin{equation}\label{tensor-relations}
\nabla_a t_{b}={\cal A}(g_{a b}-t_a t_b)\,, \quad \nabla_a  t_{b'}={\cal C}(g_{a b'}+t_a t_{b'})\,,\quad \text{and}\quad \nabla_a  g_{b c'}=-({\cal A+C})(g_{a b}t_{c'}+g_{a c'} t_{b})\,,
\end{equation}
where the scalar functions ${\cal A}$ and ${\cal C}$
are given by 
\begin{equation}\label{AC}
{\cal A}=\frac{\coth(\ell/L)}{L}\,, \qquad \text{and} \qquad {\cal C}=-\frac{\csch(\ell/L)}{L}\,.
\end{equation}
Here $L$ is the AdS radius. The functions ${\cal A}$ and ${\cal C}$ are obtained as the solutions to a set of equations that follow from \eqref{tensor-relations} and the defining relations \eqref{normals}. See  \cite{Allen:1985wd} for details.

In particular, the relevant geometry for the vector-vector correlator was developed in \cite{Allen:1985wd}.
In our work, we are interested in the particular case in which the vector is a conserved current $J_a$, whose correlator can be written as  
\begin{equation}\label{current-current-ads}
\langle J_a (x) J_{b'} (x')\rangle=-\tilde{G}(\ell)t_a t_{b'}-\tilde{F}(\ell)g_{a b'}\,.
\end{equation}
Current conservation at either point imposes a constraint on the functions $\tilde{F}(\ell)$ and $\tilde{G}(\ell)$. The relevant constraint follows from either $\nabla_x^a \langle J_a (x) J_{b'} (x')\rangle=0$ or $\nabla_{x'}^{b'} \langle J_a (x) J_{b'} (x')\rangle=0$. Evaluating the first of the above conditions, with the help of the identities \eqref{tensor-relations} and \eqref{AC}, we obtain   
 \begin{equation}
 \nabla^a\langle J_a (x) J_{b'} (x')\rangle=-\left[\tilde{G}'(\ell)+{\cal A}(d-1) \tilde{G}(\ell) -\tilde{F}'(\ell)-({\cal A+C})(d-1)\tilde{F}(\ell)\right]t_{b'}=0\,,
 \end{equation}
 which leads to a first-order differential equation relating $\tilde{F}(\ell)$ and $\tilde{G}(\ell)$. Here a prime on a function denotes differentiation with respect to its argument.  
 Substituting the expressions for the functions ${\cal A}$ and ${\cal C}$ leads to 
\begin{equation}\label{conservation-1}
 \tilde{G}'(\ell)-\tilde{F}'(\ell)=\frac{d-1}{L \sinh\left(\ell/L\right)}\left[\Big(\cosh(\ell/L)-1\Big)\tilde{F}(\ell)-\cosh(\ell/L) \tilde{G}(\ell)\right]\,.
 \end{equation}
 This equation takes a simpler form if one rewrites $\tilde{F}$ and $\tilde{G}$ as
 \begin{equation}\label{FGtildes}
 \tilde{F}(\ell)= \frac{F(\ell)}{[2L \sinh(\ell/2L)]^{2(d-1)}}\,,\quad \text{and} \quad \tilde{G}(\ell)= \frac{G(\ell)}{[2L \sinh(\ell/2L)]^{2(d-1)}}\,.
 \end{equation}
In that case, the conservation equation becomes 
\begin{equation}\label{conservation-2}
k'(\ell)=-\frac{(d-1)}{L\sinh(\ell/L)}\left[2F(\ell)-G(\ell) \right]\,, \qquad \text{where}\qquad k(\ell)\equiv G(\ell)-F(\ell)\,.
\end{equation}
For extensive theories satisfying $2F(\ell)\geq G(\ell)\geq 0$, the quantity $k(\ell)$ decreases monotonically with the geodesic length, $k'(\ell)\leq0$. If, additionally, $G(\ell)>F(\ell)$, then $k(\ell)\geq 0$ and $k(\ell)$ can therefore be interpreted as an RG monotone tracking the effective degrees of freedom coupled to the conserved current in AdS. These properties are satisfied by the correlator that comes from a free massive Dirac fermion. See the ‘Current-current correlator for a massive fermion in AdS’ section below.
 
\subsection{Correlator as a total derivative}

Solving either \eqref{conservation-1} or \eqref{conservation-2}, one can write the current-current correlator in terms of a single scalar function. However, it is geometrically more interesting to rewrite the correlator \eqref{current-current-ads} in terms of total derivatives acting on a tensor determined by a single scalar function. In particular, this facilitates the evaluation of the charge-charge correlator for arbitrary regions by applying Stokes' theorem. Our proposed ansatz is
\begin{equation}
\langle J_a (x) J^{b'} (x')\rangle=- \nabla^{\alpha} \nabla_{\beta'}\left[g\indices{_a^{[b'}}g\indices{_\alpha^{\beta']}} H(\ell)\right]\,.
\label{current-current-AdS-2}
\end{equation}
 In the flat-space limit, the ansatz reduces to $\langle J_\mu J_\nu\rangle=-(\partial_\mu\partial_\nu-\eta_{\mu\nu}\Box)H$, as in~\cite{Casini:2008wt}. Recall that $g_{\alpha}^{\,\,\, \beta'}\nabla_{\beta'} \ell=-\nabla_\alpha \ell$ since the primed derivative acts at $x'$, whereas the unprimed derivative acts at $x$. We can check that this ansatz automatically satisfies the conservation equation for any function $H(\ell)$.
This can be seen by applying the operator $\nabla^a$ to~\eqref{current-current-AdS-2}, which gives
\begin{equation}
\nabla^a\langle J_a (x) J^{b'} (x')\rangle=-\frac{1}{2} \nabla^{(a} \nabla^{\alpha)} \nabla_{\beta'}\left[g\indices{_a^{[b'}}g\indices{_{\alpha}^{\beta']}} H(\ell)\right]-\frac{1}{2}\nabla^{[a} \nabla^{\alpha]}\nabla_{\beta'}\left[g\indices{_{a}^{b'}}g\indices{_{\alpha}^{\beta']}} H(\ell)\right]\,.
\end{equation}
The contribution from the symmetrized covariant derivatives vanishes identically since the tensor on which they act is antisymmetric in $a$ and $\alpha$. The action of the antisymmetric derivative operator can be written in terms of the Riemann tensor as
\begin{align}
 \nabla_{[a} \nabla_{\alpha]}\nabla_{\beta'}\left[g^{a|[b'}g^{\alpha|\beta']} H(\ell)\right]&=-R\indices{_a_\alpha_\lambda^a}\nabla_{\beta'}\left[g^{\lambda|[b'}g^{\alpha|\beta']} H(\ell)\right]-R\indices{_a_\alpha_\lambda^\alpha}\nabla_{\beta'}\left[g^{a|[b'}g^{\lambda|\beta']} H(\ell)\right]\,,\nonumber\\
&=-2R_{a\lambda}\nabla_{\beta'}\left[g^{a|[b'}g^{\lambda|\beta']} H(\ell)\right]=0\,.
\end{align}
The above equalities follow from the definitions of the Riemann and Ricci tensors and their symmetries. Finally, it is useful to connect both expressions for the current-current correlator. In order to achieve that, we simplify the right-hand side of~\eqref{current-current-AdS-2} and put it in the form \eqref{current-current-ads}, such that $\tilde{F}(\ell)$ and $\tilde{G}(\ell)$ are expressed as functions of $H(\ell)$. Explicitly, this leads to 
\begin{align}\label{F-G-as-H}
\tilde{F}(\ell)&=-(d-2)\Big[\Big((d-2){\cal A}-{\cal C}\Big)({\cal A+C})H(\ell)+(2{\cal A}+{\cal C})H'(\ell)\Big]-H''(\ell)\,, \nonumber\\
\tilde{G}(\ell)&=-(d-2)^2({\cal A+C})^2 H(\ell)-\Big[2(d-2){\cal A}+(2d-3){\cal C} \Big]H'(\ell)-H''(\ell)\,,
\end{align}
where we used the identities~\eqref{tensor-relations}. In this way, given a function $H(\ell)$, we obtain the functions $\tilde{F}(\ell)$ and $\tilde{G}(\ell)$ that appear in the correlator~\eqref{current-current-ads}. Alternatively, given $\tilde{F}(\ell)$ and $\tilde{G}(\ell)$ that satisfy the conservation equation~\eqref{conservation-1}, we can obtain the function $H(\ell)$ by solving the first-order differential equation 
\begin{equation}
H(\ell)+\frac{L \coth[\ell/(2L)]}{d-2}H'(\ell)-\frac{2L^2\cosh^2[\ell/(2L)]}{(d-1)(d-2)}\left[\tilde{G}(\ell) -\tilde{F}(\ell)\right]=0\,,
 \end{equation}
which is obtained from~\eqref{F-G-as-H}, with~\eqref{AC}. The solution is simply 
\begin{equation}\label{H-k}
 H(\ell)=\frac{1}{\cosh^{2(d-2)}[\ell/(2L)]}\left(c_\star+\frac{1}{(2L)^{2d-3}(d-1)}\int_{\ell_\star}^\ell \frac{k(x)\diff x}{\tanh^{2d-3}[x/(2L)]}\right)\,.
 \end{equation}
Here the source term is expressed in terms of $k(x)$, defined in~\eqref{conservation-2}, using~\eqref{FGtildes}. The constant $c_\star$ is the integration constant, while $\ell_\star$ is an arbitrary reference point whose choice can be absorbed into $c_\star$.

\subsection{Correlator at the conformal fixed point}

At a conformal fixed point in flat space, the conserved-current correlator has a universal form; thus, we expect a similar universality to hold in AdS. To see this in our framework, we start with the assumption that the conformal fixed point in AdS can be characterized by the condition that the RG charge does not flow, namely, $k(\ell)=k_J$ is constant. In that case,~\eqref{current-current-ads} reduces to 
\begin{equation}\label{current-current-conformal}
\langle J_{a}(x)J_{b'}(x')\rangle = -\frac{(2t_a t_{b'}+g_{a b'})k_J}{\left(2L\sinh[\ell/(2L)]\right)^{2(d-1)}}\, ,
\end{equation}
Likewise, we obtain a closed-form expression for $H(\ell)$ via direct integration of~\eqref{H-k}, which can be written as 
\begin{equation}\label{Hl}
 H(\ell)=-\frac{ k_J}{2(d-1)(d-2)}\frac{\,_2F_1\left(1 , -(d-2), -(d-3);\tanh^2\left[\ell/(2L)\right]\right)}{\left(2L \sinh[\ell/(2L)]\right)^{2(d-2)} }\,,
 \end{equation}
 where we have set $c_\star=0$ with $\ell_\star\to \infty$. Equation~\eqref{Hl} holds for non-integer $d$; at integer $d$ the hypergeometric function has a pole proportional to the homogeneous solution $\cosh^{-2(d-2)}[\ell/(2L)]$, which is absorbed into $c_\star$. For integer $d$, $H(\ell)$ can instead be expressed in terms of elementary functions as
\begin{equation}\label{Hl2}
 H(\ell)=\frac{k_J/(d-1)}{\left(2L\cosh[\ell/(2L)]\right)^{2(d-2)}}\left\{\log\left[\sinh\left(\frac{\ell}{2L}\right)\right]-\sum_{k=1}^{d-2} \frac{\coth^{2k}\left[\ell/(2L)\right]}{2k}\right\}\,.
 \end{equation}
 In either form, the total derivative expression~\eqref{current-current-AdS-2} for the current-current correlator in AdS does not yield a simple charge-charge correlator upon integration. In our work, we identify this charge-charge correlator as providing an extensive model for mutual information (EMI).  
 
 A crucial observation is that the current-current correlator in a CFT in Minkowski half-space is related to the current-current correlator in AdS by a Weyl rescaling. This relation can be checked directly in Poincar\'e coordinates where $x=(z,{\bf{x}})$ and the geodesic length is $\ell=L\arcosh\left[1+(x-x')^2/(2zz')\right]$. Explicitly evaluating the correlator~\eqref{current-current-conformal} gives
\begin{equation}\label{JJ-AdS-Mink}
\left.\langle J_{a}(x)J_{b'}(x')\rangle\right|_\text{AdS}=(zz')^{d-2} \left.\langle J_{a}(x)J_{b'}(x')\rangle\right|_\text{Mink}\,,
 \end{equation}
 where the correlators on the left- and right-hand sides are evaluated in AdS with metric $g_{ab}$ and in Minkowski half-space with metric $\eta_{ab}$, respectively, with the Minkowski coordinates identified with the Poincar\'e coordinates. In those coordinates, the various geometric quantities in \eqref{tensor-relations} take the form
\begin{equation}\label{tensors-g-n}
g_{a b'}=-\partial_a\,  \partial_{b'} \xi+\frac{
\partial_a \xi\, 
\partial_{b'} \xi}{\xi+1}\,, \qquad t_a=\frac{\nabla_a \xi}{\sqrt{\xi^2-1}}\,,\qquad\text{and}\qquad t_{b'}=\frac{\nabla_{b'} \xi}{\sqrt{\xi^2-1}}\,,
\end{equation}
where $\xi$ is related to the geodesic distance $\ell$ via $\xi=\cosh(\ell/L)$, and to the flat-space distance $r^2\equiv(x-x')^a(x-x')^b \eta_{ab}$ via $\xi=1+r^2/(2zz')$. The tangent vectors to the geodesics in flat space are $\hat{r}_a=\partial_a r$ and $\hat{r}_{b'}=\partial_{b'} r$. In Poincar\'e coordinates, the tensorial combination that appears in~\eqref{current-current-conformal} takes the simple form 
\begin{equation}\label{id-tensor-structure}
2 t_a t_{b'}+g_{a b'}=-\partial_a \partial_{b'}\, \xi+\frac{
\partial_a \xi\, 
\partial_{b'} \xi}{\xi-1}=\frac{2 \hat{ r}_a \hat{r}_{b'}+\eta_{a b'}}{z z'}\,,
\end{equation}
which, together with the relation $2\sinh(\ell/2L)=\sqrt{2(\xi-1)}=r/\sqrt{zz'}$, establishes the equality \eqref{JJ-AdS-Mink}. Moreover, this relation implies an exact equivalence between the charge-charge correlator in AdS and that in Minkowski half-space, namely 
\begin{equation}\label{JJ-AdS=JJ-Mink}
\left.\int_X \int_Y \left[\diff \sigma_{X}\diff\sigma_Y n^{\mu}_X n^{\nu'}_Y\langle J_{\mu}(x)J_{ \nu'}(y)\rangle\right]\right|_{\text{AdS}}=\left.\int_X \int_Y \left[\diff \sigma_{X}\diff\sigma_Y n^{\mu}_X n^{\nu'}_Y\langle J_{\mu}(x)J_{ \nu'}(y)\rangle\right]\right|_{\text{Mink}}\,.
\end{equation}
Here the background labels apply to all quantities inside the brackets, including the induced measures and unit normals. Thus, although the application of Stokes' theorem in AdS to the left-hand side leads to a complicated boundary integral in terms of $H(\ell)$ \eqref{Hl}, applying Stokes' theorem in Minkowski space to the right-hand side of~\eqref{JJ-AdS=JJ-Mink} leads to the much simpler formula 
\begin{equation}\label{JJ-boundary-Mink}
-\left.k_J \int_X \int_Y\left(\diff \sigma_X\,\diff\sigma_Y\, n_X^a n_Y^{b'}\frac{(2 \hat{ r}_a \hat{r}_{b'}+\eta_{a b'})}{|x_X-x_Y|^{2(d-1)}}\right)\right|_{\text{Mink}}=\frac{k_J}{2(d-1)(d-2)}\left.\int_{\partial X}\int_{\partial Y}
          \left(\frac{\diff\sigma_X\, \diff \sigma_Y\,\mathcal{T}_{XY}}
               {|x_X-x_Y|^{2(d-2)}}\right)\right|_{\text{Mink}}\, ,
\end{equation}
where we use the explicit form of the current correlator in flat space on the left-hand side of \eqref{JJ-boundary-Mink}, and introduce the tensor 
$\mathcal{T}_{XY}\equiv\left.(n_X\cdot n_Y)(\tilde{n}_X \cdot \tilde{n}_Y) -(\tilde{n}_X \cdot n_Y)(n_X \cdot \tilde{n}_Y)\right|_\text{Mink}$,
with $n_X$ and $\tilde n_X$ denoting a pair of mutually orthogonal unit vectors normal to the codimension-two surface $\partial X$, with $n_X$ future-directed and $\tilde n_X$ pointing outward from $X$, and similarly for $Y$. We should recall that the norms and inner products are evaluated with the Minkowski metric $\eta_{ab}$. In the main text, we use the right-hand side of~\eqref{JJ-boundary-Mink} to evaluate the AdS-EMI formula at the conformal fixed point.

\section{Current-Current correlator for a Massive fermion in AdS \label{massive-fermion}}

In the main text, we use the current-current correlator of a free massive fermion to provide an explicit realization of an extensive model of MI in AdS. Such a realization leads to a family of models parametrized by $\nu=m L$ or, equivalently, by the scaling dimension of the dual operator $\Delta=(d-1)/2+\nu$. In this section, we discuss this correlator in more detail. We start with the correlator of a Dirac spinor in AdS, which can be written as 
\begin{equation}\label{ferm-corr}
S(x,x') \equiv \langle \Psi(x)\bar{\Psi}(x')\rangle
=\Big(a + b\gamma^\mu t_\mu\Big)\,\Lambda(x,x'), 
\end{equation}
where $\gamma^\mu$ are the Dirac matrices, $\Lambda(x,x')$ is the spinor parallel propagator along the geodesic connecting $x$ to $x'$, and $a$, $b$ are scalar functions of the geodesic distance $\ell$ which are fixed by the Dirac equation 
\bea\label{Dirac-eq}
\[(\gamma^\mu D_\mu-m)S(x,x')\]\indices{^\alpha_{\beta'}}=\frac{\delta(x-x')}{\sqrt{g(x)}}\delta\indices{^\alpha_{\beta'}}\,,
\eea
the short-distance flat-space divergence, and the appropriate near-boundary falloff. Explicit hypergeometric expressions for these functions are given in~\cite{Mueck:1999efk}. 
Here we present the result in terms of the distance function $\zeta:=\cosh^2(\ell/2L)$: 
\bea\label{ab-functions}
a(\zeta)&=&\lambda\, \zeta^{-\( \frac{d-1}{2}+\nu\)}\,_2F_1\(\frac{d}2+\nu , \nu, 1+2\nu;\frac1\zeta\)\,, \nonumber\\
b(\zeta)&=&\lambda \(\frac{\zeta-1}{\zeta}\)^{1/2}\zeta^{-\( \frac{d-1}{2}+\nu\)}\,_2F_1\(\frac{d}2+\nu ,1+ \nu, 1+2\nu;\frac1\zeta\) \,,
\eea
which are the only possible solutions for $\nu>1/2$. For $\nu<1/2$, there is another solution which we will not consider here. The constant $\lambda$ is given in \eqref{approxs-G-F-lambda}.
The current associated with the global $U(1)$ symmetry of the theory is given by $J_\mu(x)=\bar{\Psi}(x)\gamma_\mu\Psi(x)$ so its correlator is
\begin{equation}
\langle J_\mu(x)J_{\nu'}(x')\rangle
= -\mathrm{tr}\!\left[\gamma_\mu\,S(x,x')\,\gamma_{\nu'}\,S(x',x)\right].
\end{equation}
 Using the property $\Lambda\gamma_{b'}\Lambda^{-1}=\gamma_c g^c{}_{b'}$ of the spinor parallel propagator together with the trace identities $\mathrm{tr}(\gamma_a\gamma_c)=2^{\lfloor d/2\rfloor}\,g_{ac}$ and $\mathrm{tr}(\gamma_a\gamma_\mu \gamma_c \gamma_\nu )\,t^\mu t^\nu
= 2^{\lfloor d/2\rfloor}\,(2t_a t_c -  g_{ac})$, we obtain the following closed-form expression for the current-current correlator 
\bea
\langle J_a(x)J_{b'}(x')\rangle
=-2^{\lfloor d/2\rfloor}\[(a^2+b^2)g_{a b'}+2b^2t_a t_{b'}\]\,.
\eea
From this, we can read off the functions $\tilde{F}(\ell)$ and $\tilde{G}(\ell)$  \eqref{current-current-ads}, or equivalently, $F(\ell)$ and $G(\ell)$ \eqref{FGtildes}. These are given by
\begin{equation}
F(\ell)= \frac{2^{\lfloor d/2\rfloor}(a^2+b^2)}{\left[2L\sinh\left(\ell/2L\right)\right]^{2(1-d)}}\,, \qquad {\rm and}\qquad G(\ell)= \frac{2^{\lfloor d/2\rfloor}2b^2}{\left[2L\sinh\left(\ell/2L\right)\right]^{2(1-d)}}\,.
\end{equation}
These expressions satisfy the required positivity condition of the extensive model, namely $2F\geq G>0$, as $a^2>0$. Moreover, this correlator satisfies the condition $G\geq F$ since $b^2\geq a^2$, which implies that  
\begin{equation}
k(\ell)=2^{\lfloor d/2\rfloor} \left[2L \sinh\left(\frac{\ell}{2L}\right)\right]^{2(d-1)}(b^2-a^2)\,,
\end{equation}
is nonnegative, and therefore, it can be interpreted as a measure of the degrees of freedom coupled to the current.

\subsection{Short- and long-distance structure of the current correlator}
We need to study the short- and long-distance regimes of the current-current correlator encapsulated in the functions $F(\ell)$ and $G(\ell)$ to understand the behavior of the associated extensive model. For this analysis, it is convenient to employ the AdS-Poincar\'e coordinates $x=(z,{\bf x})$ and the associated flat-space distance function $r^2=(x-x')^a(x-x')^b \eta_{ab}$. We define the short- and long-distance limits as the regimes in which $r/L\ll 1$ and $r/L\gg 1$, respectively. At short distances, we take $r=\delta$. Thus, our distance functions $\xi$ and $\zeta$ have the following asymptotic forms
\begin{equation}
\xi= 1+\frac{\delta^2}{2z z'}, \quad \xi\approx  \frac{r^2}{2z z'}\,, \quad{\rm and}\quad \zeta=1+\frac{\delta^2}{4z z'}, \quad \zeta\approx \frac{r^2}{4z z'}\,,
\end{equation}
where we assume that $z$ and $z'$ are of order $L$. From \eqref{ab-functions}, we can write down the asymptotic form of the functions of interest. 
In the short-distance regime, we have
\begin{equation}
a(\zeta)\approx \lambda \,\frac{\Gamma(1+2\nu)\Gamma(d/2-1)}{\Gamma(d/2+\nu)\Gamma(\nu)}\frac{(4zz')^{\frac{d-2}{2}}}{\delta^{d-2}}\,, \quad{\rm and } \quad 
b(\zeta)\approx \lambda\, \frac{\Gamma(1+2\nu)\Gamma(d/2)}{\Gamma(d/2+\nu)\Gamma(1+\nu)}\frac{(4zz')^{\frac{d-1}{2}}}{\delta^{d-1}} \,.
\end{equation}
Thus, $|b(\zeta)|\gg |a(\zeta)|$ and we obtain
\begin{equation}\label{approxs-G-F-lambda}
\tilde{G}\approx 2\tilde{F}\approx 2 \lambda^2 \frac{2^{\lfloor d/2\rfloor}\,\Gamma^2(1+2\nu)\Gamma^2(d/2)}{\Gamma^2(d/2+\nu)\Gamma^2(1+\nu)}\frac{(4zz')^{d-1}}{\delta^{2(d-1)}},\quad {\rm where}\quad 
\lambda=-\frac{1}{L^{d-1}}\, \frac{2^{-(d+2\nu)}\Gamma\left(d/2+\nu\right)}{\pi^{\frac{d-1}{2}}\Gamma\left(1/2+\nu\right)}\, .
\end{equation}
Using relations \eqref{FGtildes} and $2L\sinh(\ell/2L)=L\delta/\sqrt{zz'}$, we obtain the short-distance limit of the RG charge $k(\zeta)\approx k_J$:
\begin{equation}
k_J= \frac{2^{\lfloor d/2\rfloor-2}2^{-4\nu}\,\Gamma^2(1+2\nu)\Gamma^2(d/2)}{\pi^{d-1}\Gamma^2(1/2+\nu)\Gamma^2(1+\nu)}= \frac{2^{\lfloor d/2\rfloor-2}\Gamma^2(d/2)}{\pi^d}\,,
\end{equation}
which agrees exactly with the current central charge of a Dirac spinor in $d$ dimensions $k^{\cal D}_J=2^{\lfloor d/2\rfloor-2}\Gamma^2(d/2)/\pi^d$. The above agreement was built into the normalization of the solutions to the Dirac equation \eqref{Dirac-eq}~\cite{Mueck:1999efk}.

On the other hand, in the large-distance regime, we have 
\begin{equation}
|a(\zeta)|\approx |b(\zeta)|\approx |\lambda|\, \zeta^{-\left( \frac{d-1}{2}+\nu\right)}\approx |\lambda| \frac{(4 z z')^{\Delta}}{r^{2\Delta}} \quad {\rm and} \quad \tilde{F}(\zeta)\approx \tilde{G}(\zeta)\approx 2^{\lfloor d/2\rfloor}2\lambda^2 \frac{(4 z z')^{2\Delta}}{r^{4\Delta}}\,.
\end{equation}
In the main text, we define the extensive model as proportional to the charge-charge correlator built from the above current. This takes the form 
\begin{equation}\label{mutual-SM}
\IAB = -\kappa \int_{A}\int_{B}\diff\sigma_A \, \diff\sigma_B\, \left[(n_A \cdot t) (n'_B \cdot t')\, \tilde{G}(\zeta) +(n_A \cdot n'_B)\, \tilde{F}(\zeta)\right]\, ,
\end{equation}
where the above integrals are over a Cauchy surface of AdS, and $\kappa$ is an arbitrary constant of the model. In our model we are interested in taking regions $A$ and $B$ that are the entanglement wedges of the boundary regions $A_\partial$ and $B_\partial$, respectively. In Poincar\'e coordinates, the volume elements $\diff\sigma=(L/z)^{d-1}\diff^{d-1}x$, where $\diff^{d-1}x$ is the flat-space volume element. We will also need the following geometric approximations in the long-distance regime
    \begin{equation}
    (n_A \cdot t) (n'_B \cdot t')\approx -\frac{4z  z'}{|\mathbf{x} - \mathbf{x}'|^2}(\mathbf{n}_A \cdot \mathbf{\rh})(\mathbf{n}_B \cdot \mathbf{\rh})\,\quad {\rm and}\quad 
    (n_A \cdot n'_B) \approx (\mathbf{n}_A \cdot \mathbf{n}_B)-2(\mathbf{n}_A \cdot \mathbf{\rh})(\mathbf{n}_B \cdot \mathbf{\rh})\, ,
\end{equation}
    which are obtained from the explicit expressions \eqref{tensors-g-n} with $n_A^z=n_B^z=0$. We choose the Cauchy slice to extend trivially from the RT surface to the boundary with $n^z=0$. 
    In the long-distance limit of the mutual information \eqref{mutual-SM}, we can take all factors except the powers of $z$ and $z'$ outside the integrals at leading order. Thus, using the above approximations we obtain
 \begin{equation}
I(A,B)\sim 
\frac{\kappa\, k_\infty \, C_A C_B\left[2({\bf n}_A \cdot {\bf \hat{r}}) ({\bf n}_B \cdot {\bf \hat{r}})-{\bf n}_A \cdot {\bf n}_B\right]}{|\mathbf{x}_A-\mathbf{x}_B|^{4\Delta}} \,,
    \end{equation}
where ${\bf x}_A$ and ${\bf x}_B$ are any pair of chosen points at $A_\partial$ and $B_\partial$, respectively. The coefficients $C_X$ with $X=A,B$ depend on the shape of the entangling regions, while the constant $k_\infty$ comes from the normalization of the current-current correlator. These are explicitly given by
\begin{equation}
C_X=\int_{X} \diff^{d-1}x_X z_X^{2\Delta-(d-1)}\,,\qquad {\rm and}\qquad k_\infty=\frac{2^{\lfloor d/2\rfloor-1}\Gamma^2(d/2+\nu)}{\pi^{d-1}\Gamma^2(1/2+\nu)}\,.
\end{equation}
At the conformal fixed point, $k(\ell)=k_J$ for all $\ell$, and the long-distance coefficient $k_\infty=k_J$. On the other hand, taking the $\nu\to 0$ limit of the above expression leads to the relation $k_{\infty}(\nu=0)=2k^{\cal D}_J$. This is due to the fact that this solution does not interpolate smoothly between the massive and massless cases. Consequently, the conformal fixed point cannot be obtained by taking the $\nu\to0$ limit of this solution.
\end{document}